\documentclass[aps, prb, superscriptaddress, preprint]{revtex4-1}

\usepackage{graphicx}% Include figure files
\usepackage{color}
\usepackage{dcolumn}% Align table columns on decimal point
\usepackage{bm}% bold math
\begin{document}

%\preprint{APS/123-QED}

\title{Energy scale of nematic ordering in the parent iron-based superconductor:BaFe$_2$As$_2$}% Force line breaks with \\

\author{Alexander Fedorov}
 \affiliation{IFW Dresden, P.O. Box 270116, D-01171 Dresden, Germany}

\author{Alexander Yaresko}
\affiliation{Max Planck Institute for Solid State Research, Heisenbergstrasse 1, 70569 Stuttgart, Germany}
\author{Erik Haubold}
 \affiliation{IFW Dresden, P.O. Box 270116, D-01171 Dresden, Germany}
\author{Yevhen Kushnirenko}
 \affiliation{IFW Dresden, P.O. Box 270116, D-01171 Dresden, Germany}
\author{Timur Kim}
 \affiliation{Diamond Light Source, Harwell Campus, Didcot OX11 0DE, United Kingdom}
\author{Bernd B\"uchner}
 \affiliation{IFW Dresden, P.O. Box 270116, D-01171 Dresden, Germany}
\author{Saicharan Aswartham}
 \affiliation{IFW Dresden, P.O. Box 270116, D-01171 Dresden, Germany}
\author{Sabine Wurmehl}
 \affiliation{IFW Dresden, P.O. Box 270116, D-01171 Dresden, Germany}
\author{Sergey Borisenko}
 \affiliation{IFW Dresden, P.O. Box 270116, D-01171 Dresden, Germany}

\date{\today}% It is always \today, today,

\begin{abstract}

Nematicity plays an important role in the physics of iron-based superconductors (IBS). Its microscopic origin and in particular its importance for the mechanism of high-temperature superconductivity itself are highly debated. A crucial knowledge in this regard is the degree to which the nematic order influences the electronic structure of these materials. Earlier angle-resolved photoemission spectroscopy (ARPES) studies found that the effect is dramatic in three families of IBS including 11, 111 and 122 compounds: energy splitting reaches 70 meV and Fermi surface becomes noticeably distorted. More recent experiments, however, reported significantly lower energy scale in 11 and 111 families, thus questioning the degree and universality of the impact of nematicity on the electronic structure of IBS. Here we revisit the electronic structure of undoped parent BaFe$_{2}$As$_{2}$ (122 family). Our systematic ARPES study including the detailed temperature and photon energy dependencies points to the significantly smaller energy scale also in this family of materials, thus establishing the universal scale of this phenomenon in IBS. Our results form a necessary quantitative basis for theories of high-temperature superconductivity focused on the nematicity.

\begin{description}

\item[PACS numbers]

\end{description}

\end{abstract}

\maketitle

\section{Introduction}

Nematic order in iron-based superconductors (IBS) is a robust experimental fact and seems to be an essential ingredient of the superconductivity~\cite{Fernandes2014,Li2017,Chubukov_PhT,Kreisel_2017}. The transition from tetragonal to orthorhombic phase is of electronic origin~\cite{Fernandes2014,Li2017,Fedorov2016,Chubukov_PhT,Kreisel_2017, Watson2015_1} and it is highly debated whether the striped magnetic or orbital ordering is directly responsible for this phenomenon~\cite{Chubukov_PhysRevX.6.041045,Baek2014}.

 In order to advance in this debate it is necessary to provide quantitative theories with exact characterization of electronic nematicity. One of the quantitative estimates in terms of energy and momentum comes from angle-resolved photoemission~\cite{Yi6878,Shimojima2014, Nakayama2014,Watson2015_1,Zhang2015,Fanfarillo2016,Xu2016,Ye2015,MYiNJP2012}. In BaFe$_{2}$As$_{2}$ the energy splitting was reported reaching 70 meV~\cite{Yi6878}, in NaFeAs - 40 meV~\cite{MYiNJP2012,ZhangPRB12}, in FeSe - 60 meV~\cite{Yi6878,Shimojima2014, Nakayama2014,Watson2015_1,Zhang2015,Fanfarillo2016,Xu2016,Ye2015} . Such a strong modification of the electronic structure could reveal the dominant interactions which are able to drive the pairing in IBS. However, recent re-visits of the electronic structure of some of the main stoichiometric members of the IBS family, FeSe and NaFeAs, demonstrated that the energy scale corresponding to the nematic ordering has to be re-evaluated ~\cite{Fedorov2016,Watson2016,WatsonNaFeAsPRB18}. The energy scale appears to be significantly smaller than it was believed earlier, of the order of 10-15 meV. Since the very first evidence came from the ARPES measurements of the archetypal BaFe$_{2}$As$_{2}$ (Ba122) systems~\cite{MYiPRB2009, Yi6878}, it is highly desirable to establish the energy and momentum scale also in this key family of the IBS materials. Moreover previous ARPES studies of Ba122~\cite{PhysRevLett.104.057002,MYiPRB2009,PhysRevB.81.060507,NAKASHIMA201316,Zabolotnyy2009} did not converge to a common picture. The discrepancies are mostly related to three-dimensionality, nesting conditions as well as disagreement with the band-structure calculations and bulk sensitive dHvA experiments~\cite{PhysRevB.80.064507,PhysRevLett.107.176402} . Several experiments have been carried out on the detwinned samples~\cite{Yi6878,PhysRevB.83.064509}. Again, the conclusions drawn in these studies are controversial. Kim et al.~\cite{PhysRevB.83.064509} found a good agreement with the band structure calculations, whereas Yi et al.~\cite{Yi6878} detected an unbalanced occupation of the $d_{xz}$ and $d_{yz}$ orbitals which develops fully at the transition temperature.

In this study we re-visit the electronic structure of the stoichiometric parent Ba122 compound.  We use high-resolution ARPES and conventional band-structure calculations in order to understand the fine details of the low-energy electron dynamics and its evolution in a broad temperature interval. A step-by-step analysis of the influence of the three-dimensionality, nematic and spin-density wave (SDW) transitions on the electronic structure of Ba122 allowed us to single out the optimal conditions for the experiment which directly provides the quantitative estimate of the energy and momentum scales related to the nematicity in this basic but important material.

\section{Experimental and computational details}

\subsection{Experimental details}
High quality single crystals of  BaFe$_2$As$_2$ were grown by the self  flux technique and were characterized by several complementary methods as described in Ref.~\onlinecite{ASWARTHAM_BFA}. Ba122 exhibits stripe-type anteferromagnetic ordering of Fe spins below the Neel temperature $\approx$140 K, accompanied by a structural phase transition. The high temperature  paramagnetic  tetragonal structure has lattice parameters $a=b=3.9625\AA$ , $c=13.0168\AA$ and while orthorombic phase has $a=5.6146\AA$ ,$b=5.5742\AA$, $c=12.9453\AA$~(Ref. \onlinecite{PhysRevB.78.020503}). Neutron diffraction studies have shown that the magnetic modulation vector is $Q=(1,0,1)$  in the above orthorhombic setting, with the spin orientation parallel to $a$~(Ref.\onlinecite{PhysRevLett.101.257003}).

ARPES measurements were performed at the I05 beamline of Diamond Light Source, UK\cite{Hoesch_beamline}. Single-crystal samples were cleaved \textit{in situ} in a vacuum lower than 2$\times$10$^{-10}$ mbar and measured at temperatures ranging from 5.7 to 270 K. Measurements were performed using linear horizontal (LH) and linear vertical (LV) polarized synchrotron light with variable photon energy, utilizing Scienta R4000 hemispherical electron energy analyzer with an angular resolution  of 0.1 deg and an energy resolution of 3 meV.

\subsection{Computational details}

Self-consistent band structure calculations were performed using the linear
muffin-tin orbital (LMTO) method \cite{And75} in the atomic sphere
approximation (ASA) as implemented in PY LMTO computer code
\cite{book:AHY04}. The Perdew-Wang parametrization \cite{PW92} was used to
construct the exchange correlation potential in the local density
approximation (LDA). Spin-orbit coupling (SOC) was taken into account at the
variational step.

\section{Results and discussion}

%Previous ARPES studies of Ba122~\cite{PhysRevLett.104.057002,MYiPRB2009,PhysRevB.81.060507,NAKASHIMA201316,Zabolotnyy2009} ( LiuPRL102 16704) did not converge to a common picture. The discrepancies are mostly related to three-dimensionality, nesting conditions as well as disagreement with the band-structure calculations and bulk sensitive dHvA experiments~\cite{PhysRevB.80.064507,PhysRevLett.107.176402} . Several experiments have been carried out on the detwinned samples~\cite{Yi6878,PhysRevB.83.064509}. Again, the conclusions drawn in these studies are controversial. Kim et al.~\cite{PhysRevB.83.064509} found a good agreement with the band structure calculations, whereas Yi et al.~\cite{Yi6878} detected an unbalanced occupation of the $d_{xz}$ and $d_{yz}$ orbitals which develops fully at the transition temperature.

\begin{figure*}[t]
	\centering
		\includegraphics[width=16cm]{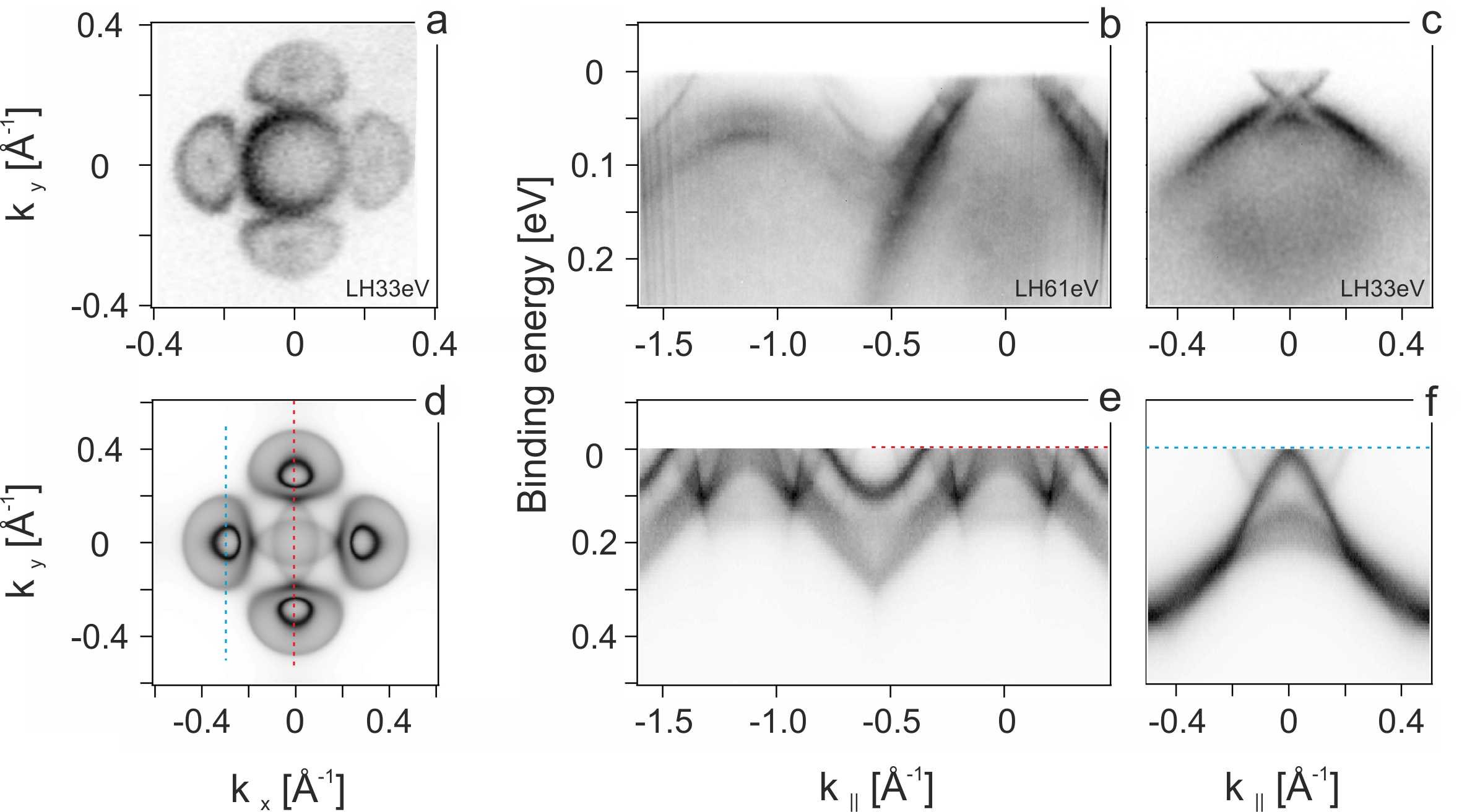}
	\caption{ARPES data accompanied by corresponding DFT calculated bandstucture in SDW phase without SOC and fully integrated over $\vec{k_z}$ in presence of two rotational domains: a),d) Fermi surface in SDW phase. b),e) Cut along the $\Gamma-M$ direction. c) f) Cut along the cyan dashed line. Only those bands which cross the $E_f$ are shown in DFT data.  }
	\label{Fig:Intro}
\end{figure*}

Taking into account all previous measurements and controversies, we first present the data recorded at low temperatures, deep in the magnetic phase and compare them with the corresponding band structure calculations integrated along $\vec{k_z}$ in Fig.1. A striking agreement between the projection of the theoretical Fermi surface on $\vec{k_x}-\vec{k_y}$ plane and experimental Fermi surface map in Fig.1 a, d has never been observed earlier. The progress in the data quality could be best seen by comparison of this map with our previous study~\cite{Zabolotnyy2009}, thanks to improvement of the beam characteristics and sample positioning at low temperatures. All features seen in the experiment are qualitatively reproduced by the band structure calculations, namely the big hole-like structure in the center, four big electron-like pockets and four hole-like small pockets which appear to be inside the electron-like ones. Moreover, the size and the topology of the measured Fermi surface now agrees with the dHvA measurements~\cite{PhysRevLett.107.176402} taking into account that we here observed signal from two rotational domains simultaneously. %\textcolor[rgb]{0,0,1}{HERE ONE CAN COMPARE THE FS AREAS WITH THE FREQUENCIES OF TERASHIMA.} 

%\tableofcontents
\begin{figure*}[t]
	\centering
		\includegraphics[width=16cm]{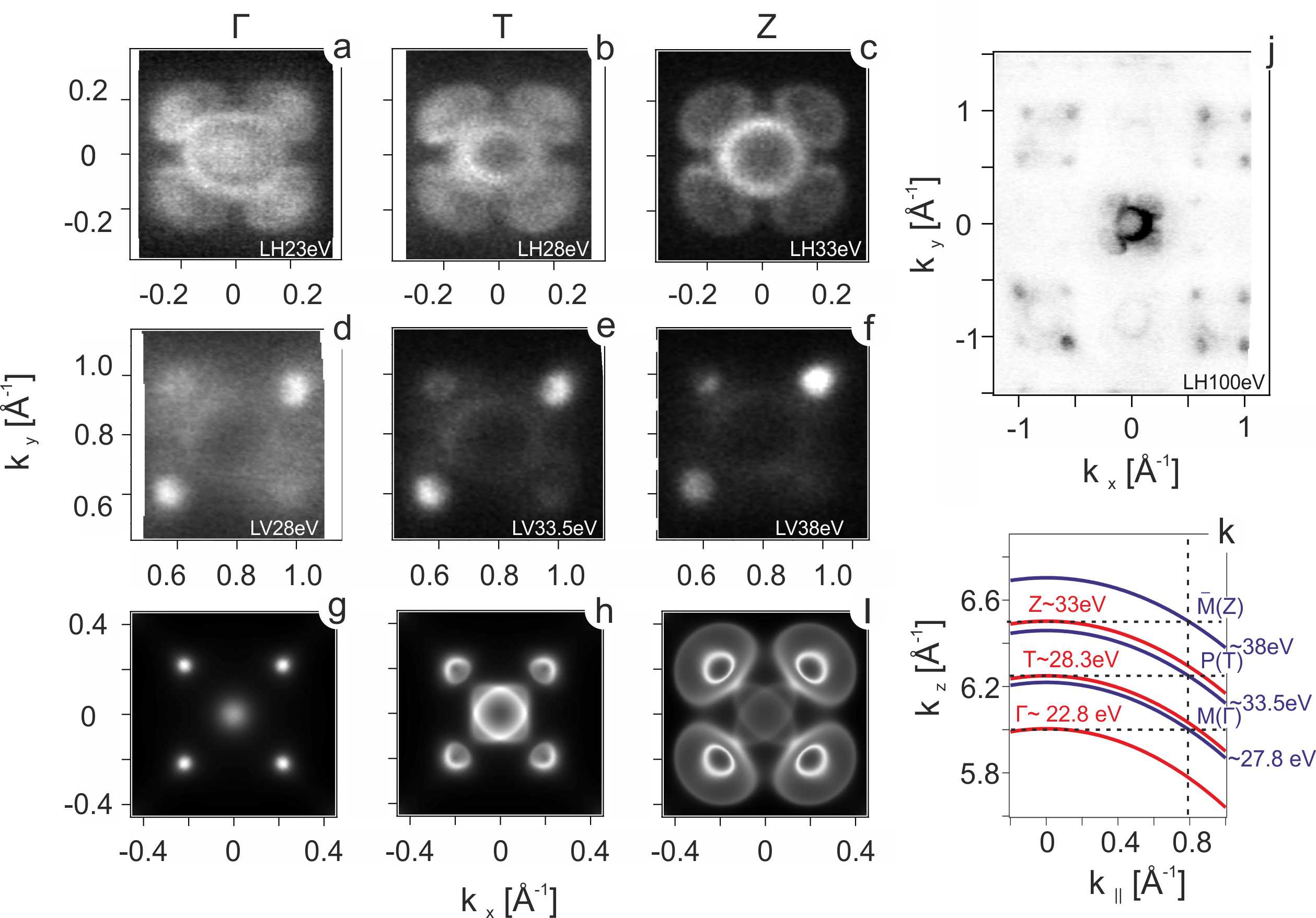}
	\caption{ a)-f) ARPES derived Fermi surfaces recorded at different photon energies corresponding to the high symmetry point in the BZ. g)- i) Calculated Fermi surfaces in the corresponding high symmetry points of the BZ integrated over the quarter in $\vec{k_z}$ direction. j) overview FS map recorded at 100eV photon energy. k) $\vec{k_z}$ vs $\vec{k_{||}}$ relations }
	\label{Fig:Kz}
\end{figure*}

 The underlying dispersions are compared in Fig.1. Also here one can notice a remarkable correspondence of the calculated dispersions with the experimental ones. Basically, all the states predicted by the mean field theory can be identified in experimental data, at least on a qualitative level. In order to achieve a quantitative agreement, one needs to take into account several known modifications. These are the global bandwidth renormalization, the relative energy shift between hole- and electron-pockets in the tetragonal state (blue-red shifts)~\cite{Borisenko2015, Fedorov2016, PhysRevB.96.100504} and matrix element effects which govern the intensities of the particular features.  Hole-like dispersions in the center of the BZ support large hole-like FS pockets which appear smaller in theory (Fig. 1b). Taking latter into account we tuned the theoretical Fermi. The shift down of 50 meV  brought the intense "V"-like crossings (at ~50-70 meV) closer to the Fermi level, as in the experiment. The  controversial doublet feature at 50-70 meV binding energy~\cite{PhysRevB.83.064509} is now also reproduced by the calculations. The size of the four small hole-like pockets can be adjusted by 50 meV upshift  of the Fermi level which is shown in Fig.1f. Obviously, because of the blue-red shifts still seen in the tetragonal phase, a single rigid shift of the band structure would not be sufficient in a folded magnetic state in order to reproduce  features related originally to hole and electron pocket simultaneously.

Based on nearly good correspondence between the experiment and the band structure calculations in Fig.1  two questions rise: (i) is the full integration over $\vec{k_z}$ essential and (ii) where is the strong influence of the nematic energy scale of the order of 70 meV ? To answer the first question, we have carried out extensive ARPES experiments to study the $\vec{k_z}$-dependence of the electronic structure. The usual way to perform such a study is to vary photon energy of excitation light thus full momentum $\vec{k}$ of emitted electron becomes controlled. Notably, using a vacuum UV photons suitable for high resolution ARPES leads to relatively small $\vec{k}$ values. Therefore $\vec{k_z}$ projection becomes very sensitive to an emission angle i.e. corresponding  $\vec{k_{||}}$. We demonstrate this fact in Fig. 2 k together with optimized photon energy values corresponding to different high symmetry points in the BZ. We have identified these points considering extensive photon energy dependent measurements which are in a good agreement with previously reported values~\cite{PhysRevB.81.060507} (See SFig.1).

We present the data in the SDW regime taken in different high symmetry points of the BZ in the center of each shown map in Fig.2a-f as well as the large overview FS map in Fig. 2g  recorded at 100 eV photon energy. In the mean field theory, the maps in the upper row of experimental data should be equivalent to the ones from the lower row. However, it is known that even without influence of the matrix elements, the photocurrent strongly depends on the degree of the potential modulations in the density wave state~\cite{Zabolotnyy_EPL2009}. That is why the intensity distribution in the corner of tetragonal BZ at all $\vec{k_z}$ values are different from the ones in the center of the BZ. In particular, the four smallest Fermi surfaces are always brighter in the second row of panels while the central rounded ones and the larger deformed ellipses adjacent to it are more clearly visible in the upper row. In panels Fig. 2 h-j we show for comparison the theoretical cuts through the FS now integrated in the smaller $\vec{k_z}$-interval (0.25 BZ). As it is seen from Fig.2 while the general $\vec{k_z}$-evolution can be tracked in the experimental data, there is always a certain admixture of the signal from different $\vec{k_z}$ values. This indicates that the $\vec{k_z}$-selectivity of this particular experiment is not very high, probably implying the smaller electron escape depth and thus lower $\vec{k_z}$-resolution. Practically, however, the intensity distribution changes when switching from map to map, indicating the possibility of finding the conditions, optimal for observation of only desired  $\vec{k_z}$-portion of the BZ. For instance, the maps centered at non-zero momentum do not have a strong signal contribution from the big hole-like pocket. This is because the modulating potential of SDW is not strong enough to make the corner and the center of the tetragonal BZ equivalent. As in the case of FeSe~\cite{Fedorov2016,Watson2016} and NaFeAs~\cite{WatsonNaFeAsPRB18} , the calculations suggest that the most convenient set of features to determine the energy scale of nematic order are the dispersions which support the FS in the corner of the tetragonal BZ. However, because Ba122 has a double-layer structure, there is an important difference overlooked earlier.
 
 \begin{figure}
	\centering
		\includegraphics[width=8cm]{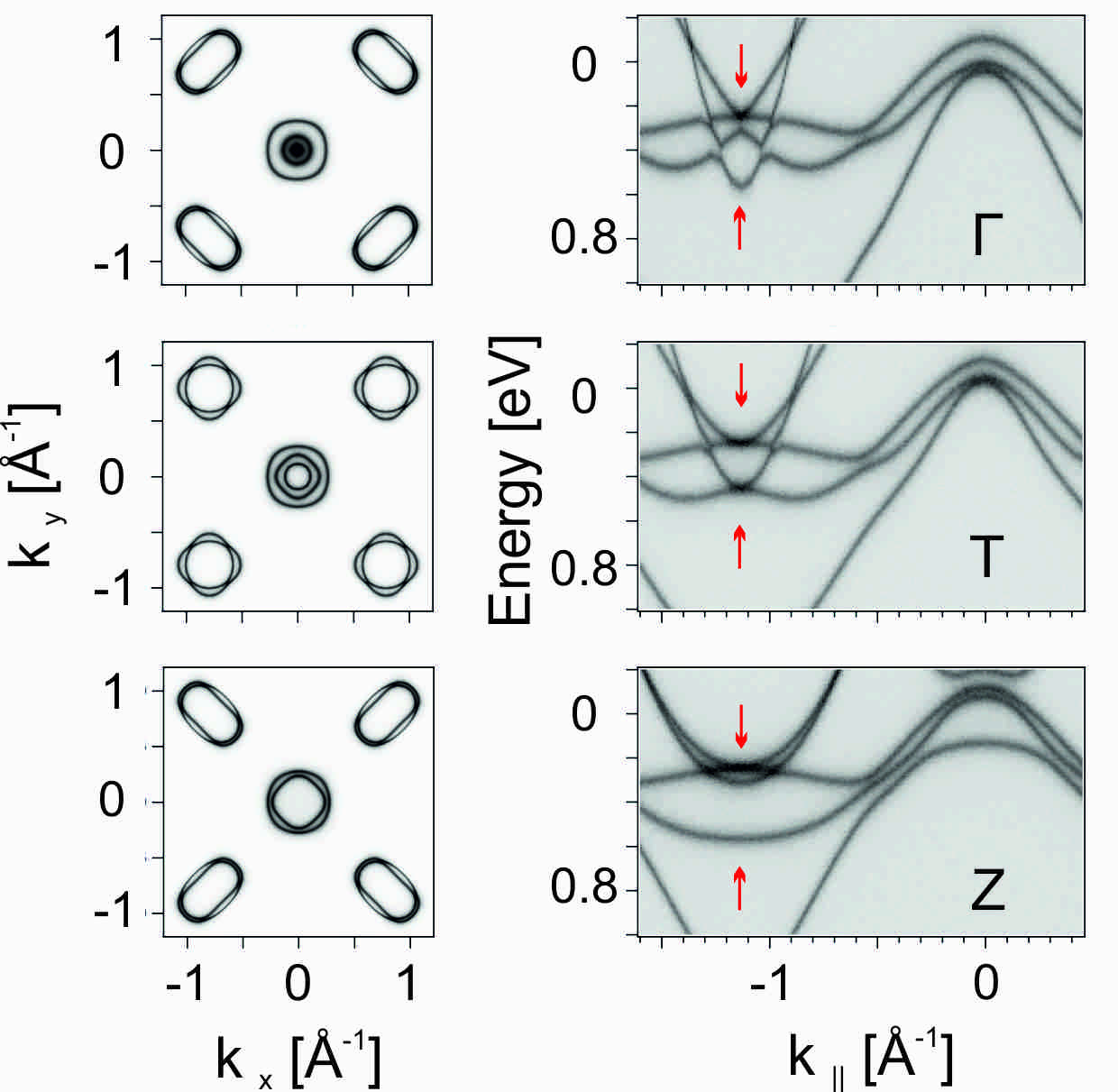}
	\caption{Band structure calculated in the tetragonal phase for three high symmetry planes in the BZ. }
	\label{Fig:Nematic}
\end{figure}

In Fig. 3 we show the results of the band structure calculations in the tetragonal phase. First of all, band structure calculations imply significant $\vec{k_z}$-dispersion and this fact alone actually prohibits the comparison of the data taken using the single photon energy along the diagonal cut with the calculated result for constant $\vec{k_z}$-value: going away from the center of the BZ, the probed $\vec{k_z}$-values will constantly decrease and upon reaching the corner of the BZ will correspond to completely different value (see Fig.2k). The most convenient place to track the influence of the nematicity on the electronic structure in Ba122 is at the corner of the BZ where the electron-like dispersions have their bottoms and spin-orbit splitting is zero~\cite{FernandesVafek}. The tops of hole-like bands are not occupied and there is admixture of the spin-orbit splitting~\cite{Borisenko2015}. However, the data presented in Fig.3 clearly show that the most suitable for this purpose is the $TNP$ plane where the energy distribution is expected to have only two peaks, closely following the dispersion in 11 and 111 compounds~\cite{WatsonNaFeAsPRB18,Fedorov2016,Watson2016}.
  
	\begin{figure*}
	\centering
		\includegraphics[width=11cm]{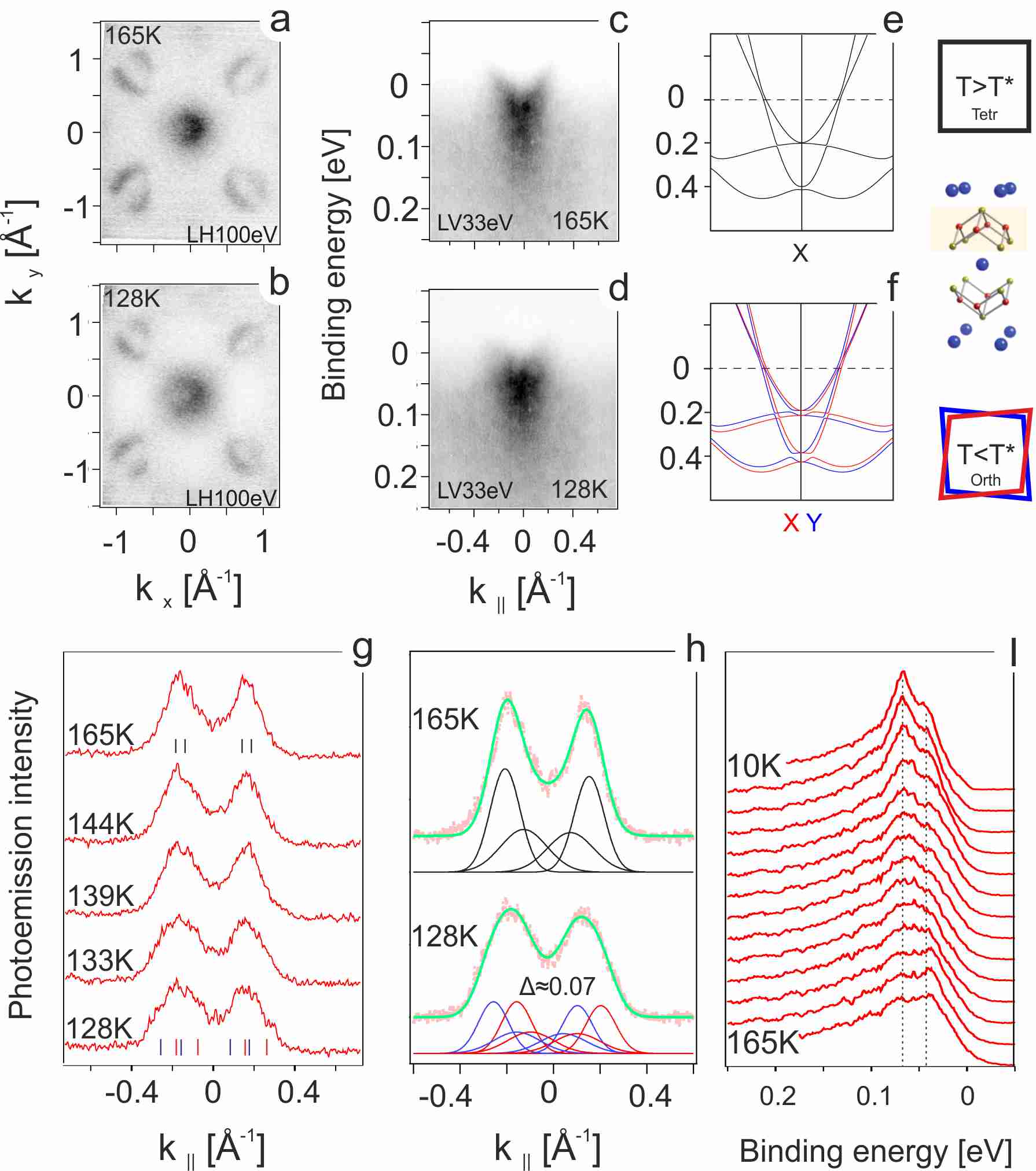}
	\caption{a),b) Overview ARPES maps recorder at 100eV photon energy above and below Tc. c),d) High resolution ARPES data recorder in the T--N direction above and below Tc. e),f) corresponding DFT calculations in tetragonal anf orthorombic phases. g) Temperature evolution of MDC cut taken on the Fermi energy corresponding to panel c). h) Fit results of MDC curves above and below Tc. i) Temperature evolution of EDC curve taken in the corner of the BZ. }
	\label{Fig:Folding}
\end{figure*}

 Now we can switch to high-temperature measurements and see what exactly happens at the transition in Fig.4. We start with overview maps recorded with 100 eV photon energy slightly above and below transition. From the first glance we can not identify any significant changes between corresponding maps and traces of the folded band structure visible at low temperature map in Fig.2. Thus we can conclude that just below transition the influence of the magnetic modulations is still weak in ARPES and most changes corresponds to the accompanying structural transition. Now we turn to the underlying dispersions of electron pocket. In analogy with FeSe~\cite{Fedorov2016} we performed diagonal T--N cuts (Fig.4) and corresponding DFT calculations for tetragonal phase (Fig.4) and  orthorhombic phase with two domains (Fig.4). Similarly, we expect to observe evolution in the corresponding EDC in the corner of the BZ from two  to four features. However, experimental observation of the four features structure is difficult even down to the 10~K (Fig.4) because of higher broadening caused by a self energy and moderate resolution in $\vec{k_z}$ direction. Indeed we always observe two peaks separated by 70 meV. Previously these two states were not resolvable at high temperature~\cite{Yi6878} and evolution from the single to double feature was assigned to the nematic transition. Here we observe that these states are an essential component of the band structure, which is now fully in agreement with corresponding DFT calculations.  

Since analysis of EDC is complicated  we switched to corresponding  MDC cuts at the Fermi energy. In tetragonal phase we expect four peaks corresponding to two electron pockets. Below orthorhombic transition and in the presence of two domains we expect twice as many features. In experimental data above transition temperature we observe MDC which can be well fitted by four lorentzian peaks (Fig.4). Slightly below the transition, we observe significant broadening of corresponding MDC. We used two sets of four peaks with same parameters as in tetragonal case to describe the low-temperature MDC. From our fit we estimate the magnitude of nematic splitting in momentum of about 0.07 \AA$^{-1}$. Using an average dispersion relation (e.i. Fermi velocity) about 0.3 eV/\AA$^{-1}$ we estimate nematic energy scale to be of the order of 20 meV.

  The difficulty to determine the degree to which nematic order changes the electronic structure in Ba122 is that it sets in simultaneously with the SDW order. This is because the magnetic order folds the bands and opens the significant interaction gaps. However, our data presented above suggest that the particular experimental conditions allow to see the portions of the momentum space where the folded replica are weak and the consequences of the structural transition can be estimated separately. In analogy with FeSe and NaFeAs, one can do this at the corner of the tetragonal BZ watching the bottoms of the electron-like pockets with the important difference that this can be done only for particular $\vec{k_z}$-interval, namely near the P point of the BZ. As we have shown above this is not straightforward since a considerable admixture of the signal from other $\vec{k_z}$ values is observed. This latter observation explains why the previous studies did not reach a consensus as regards the exact $\vec{k_z}$-dependence. Moreover, we have not detected a significant unbalanced occupation of the $d_{xz}$ and $d_{yz}$ orbitals at the transition temperature and nearly 40 K below it. At still lower temperatures the folded SDW replicas prohibit further tracking of the features related to nematic order alone. In principle, extracted by us values of nematic energy and momentum scales can already be concluded from the fact that the conventional magnetic calculations fully describe the electronic structure observed experimentally at low temperatures. Our results thus establish the universality of the strength of nematic ordering in the iron-based superconductors and provide the quantitative basis for the theories of high-temperature superconductivity in iron-based materials.

\section{Conclusions}

We re-visited the electronic structure of the stoichiometric parent Ba122 using high resolution ARPES and conventional band-structure calculations. We demonstrated that general features of the band structure in the SDW phase are well described by magnetic DFT calculations. Our calculations reveal significant $\vec{k_z}$-dispersion in this compound and experimental data show moderate $\vec{k_z}$ resolution causing additional energy and momentum broadening. Optimizing experimental conditions we were able to detect band structure modification crossing the critical temperature and quantitative estimate the energy scale of 20 meV related to the nematicity.

\section*{Acknowledgments}

We are grateful to Andrey Chubukov, Matthew Watson, Peter Hirschfeld, Roser Valenti, Sahana Roessler, Ilya Eremin and Seung-Ho Baek for the fruitful discussions. 
This work has been supported by Deutsche Forschungsgemeinschaft  DFG through the Priority Programme SPP1458
(Grant No. BU887/15-1), through grantDFG-GRK1621, and through the Emmy Noether Programme WU595/3-3 (S.W.).
We acknowledge Diamond Light Source for access to beamline I05 (proposals no.  SI13856-1 and SI15936-1) that contributed to the results presented here.

\section*{Author contributions statement}

A.F., T.K., Y.K., E. H. and S. B.  conceived the experiments. DFT calculations were performed by A.Y. The high quality BaFe$_{2}$As$_{2}$ crystals were grown by S.A. ans  S.W.  The data were discussed with B.B.  Manuscript was written by A.F and S.B.  All authors have read and approved the decisive version of the manuscript.

\section*{Additional information}

\textbf{Competing financial interests} The authors declare no competing financial interests.

%\bibliography{paperbase}


\end{document}